\documentclass[prb,amsmath,amssymb,floatfix,twocolumn,citeautoscript]{revtex4}
\usepackage{bm}
\usepackage{epsfig}
\usepackage{simplewick}
\usepackage{subfigure}
\usepackage{lipsum}
\usepackage[usenames]{color}

\begin{document}

\title{Strange Metals from Quantum Geometric Fluctuations of Interfaces}

\author{Jian-Huang She$^1$, A. R. Bishop$^{2}$, Alexander V. Balatsky$^{3, 4}$}

\affiliation{$^1$Department of Physics, Cornell University, Ithaca, New York 14853, USA.\\
$^2$Directorate for Science, Technology, and Engineering, Los Alamos National Laboratory, Los Alamos, NM, 87545, USA.\\
$^3$Institute for Materials Science, Los Alamos National Laboratory, Los Alamos, New Mexico 87545, USA.\\
$^4$Nordic Institute for Theoretical Physics (NORDITA), Center for Quantum Materials, Roslagstullsbacken 23, S-106 91 Stockholm, Sweden}

\begin{abstract}

Our current understanding of strongly correlated electron systems is based on a homogeneous framework. Here we take a  step going beyond this paradigm by incorporating inhomogeneity from the beginning. Specifying to systems near the Mott metal-insulator transition, we propose a real space picture of itinerant electrons functioning in the fluctuating geometries bounded by interfaces between metallic and insulating regions. In 2+1-dimensions, the interfaces are closed bosonic strings, and we have a system of strings coupled to itinerant electrons. When the interface tension vanishes, the geometric fluctuations become critical, which gives rise to non-Fermi liquid behavior for the itinerant electrons. In particular, the poles of the fermion Green's function can be converted to zeros, indicating the absence of propagating quasiparticles. Furthermore, the quantum geometric fluctuations mediate Cooper pairing among the itinerant electrons, indicating the intrinsic instability of electronic systems near the Mott transition.

\end{abstract}

\date{\today \ [file: \jobname]}

 \maketitle

\section{Introduction}  

\begin{figure}[t]
\begin{centering}
\includegraphics[width=0.8\linewidth]{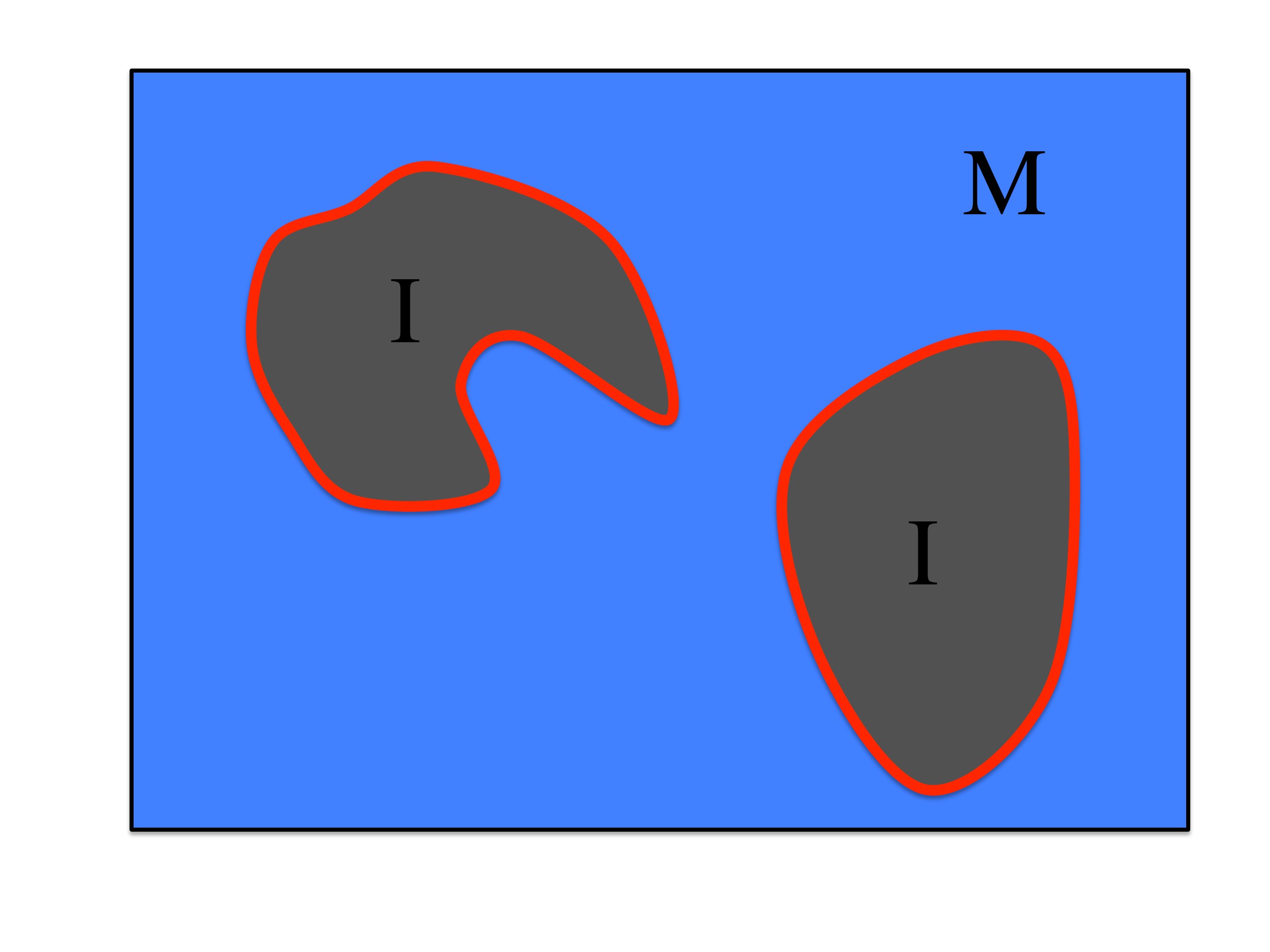} 
\end{centering}
\caption{(Color online) Real space configuration of coexisting metallic (M) and insulating (I) regions, separated by the interfaces.}
\label{Fig:real}
\end{figure}

The emerging picture of strongly correlated electron systems is that they possess a multiplicity of nearly degenerate ground states [\onlinecite{Laughlin00, Sachdev, Fradkin15}]. These systems are qualitatively different from those with a single stable ground state, like Landau Fermi liquids and symmetry broken phases. A small change in the external parameters, e.g. pressure, doping, magnetic field, substantially alters the macroscopic properties of these systems. A grand challenge of modern condensed matter physics is to build a theoretical framework for such systems with large ground state degeneracy. When the excitation energy is higher than the energy barrier between different ground states, the system is intrinsically inhomogeneous, with different ground states coexisting and fluctuating in real space [\onlinecite{Emery93, Laughlin00, Schmalian05, She10}]. The current paradigm of many body systems based on a homogeneous framework, with inhomogeneities treated as perturbations, is not adequate for the description of such intrinsically inhomogeneous systems. Here, we propose a framework that explicitly incorporates ground state degeneracy and inhomogeneity from the begining. The basis of our formalism is the quantum geometric fluctuations associated with interfaces between different degenerate ground states. The study of emergent phenomena at interfaces has recently attracted much attention [\onlinecite{Ohtomo04, Reyren07, Bozovic11, Hwang12}], which provides further  motivations for our work.

For concreteness, we consider the Mott metal-insulator transition (MIT) [\onlinecite{Mott74, Imada98}], which is prototypical of strongly correlated electron systems. At the MIT, the metallic phase and the insulating phase are degenerate in free energy. In a coarse-grained picture, each coarse-grained region can be treated as in either the metallic or the insulating state (see Fig.\ref{Fig:real}). 
 Let us focus on the charge degrees of freedom. Then the insulating regions can be treated as inert, e.g. not contributing to transport or Cooper pairing. The metallic regions are the active degrees of freedom. A crucial observation is that the metallic regions are bounded by fluctuating boundaries, and these boundaries are themselves active degrees of freedom. Then our basic picture of a Mott MIT consists of regions of itinerant electrons coupled to the fluctuating boundaries. Such a picture is fundamentally different from usual condensed matter systems, which are generally defined on a rigid background. By contrast, in our formalism, the electronic system near the MIT constitutes a physical system operating on a soft background with quantum mechanically fluctuating geometry and topology. Thermally fluctuating geometries, which we draw analogy with at various points, have been widely explored in soft condensed matter physics [\onlinecite{NelsonPiranWeinberg, Chaikin, Helfrich73, Huse88, Golubovic90, Roux92}]. However, the quantum counterpart, to the best of our knowledge, was previously encounted only in quantum gravity [\onlinecite{Polchinski}]. The closest example in condensed matter physics is probably the idea of fluctuating stripes [\onlinecite{Zaanen01}].

Such quantum geometric fluctuations have far-reaching implications for the response of a system near a second-order or weakly first-order MIT, for which the geometric fluctuations become (nearly) critical. The relevant physical systems include cuprates [\onlinecite{Imada98, Lee06}], organic conductors [\onlinecite{Powell11, Ardavan12}], iron pnictides [\onlinecite{Paglione10, Shibauchi14}] and heavy fermions [\onlinecite{Stewart01, Lohneysen07}]. These systems display in their phase diagram a metallic region with anomalous single-particle and transport properties vastly different from that of Landau Fermi liquids. Such strange metal or non-Fermi liquid phases are usually associated with abrupt changes of the Fermi surface, which signals an itinerant to localized transition for certain degrees of freedom. Strange metal phases are believed by many to be key to the understanding of high temperature superconductivity [\onlinecite{Anderson87, Lee06, Zaanen11}]. Indeed we find that the geometric fluctuations of the interface lead to anomalous scaling in the fermion self-energy. Furthermore, due to inversion symmetry breaking around the interface, couplings not allowed in a homogeneous environment are enabled [\onlinecite{Haraldsen11, Haraldsen12}], which induce attractive interactions among the fermions and promote pairing.

\section{Global phase diagram} 

\begin{figure}
\begin{centering}
\includegraphics[width=0.9\linewidth]{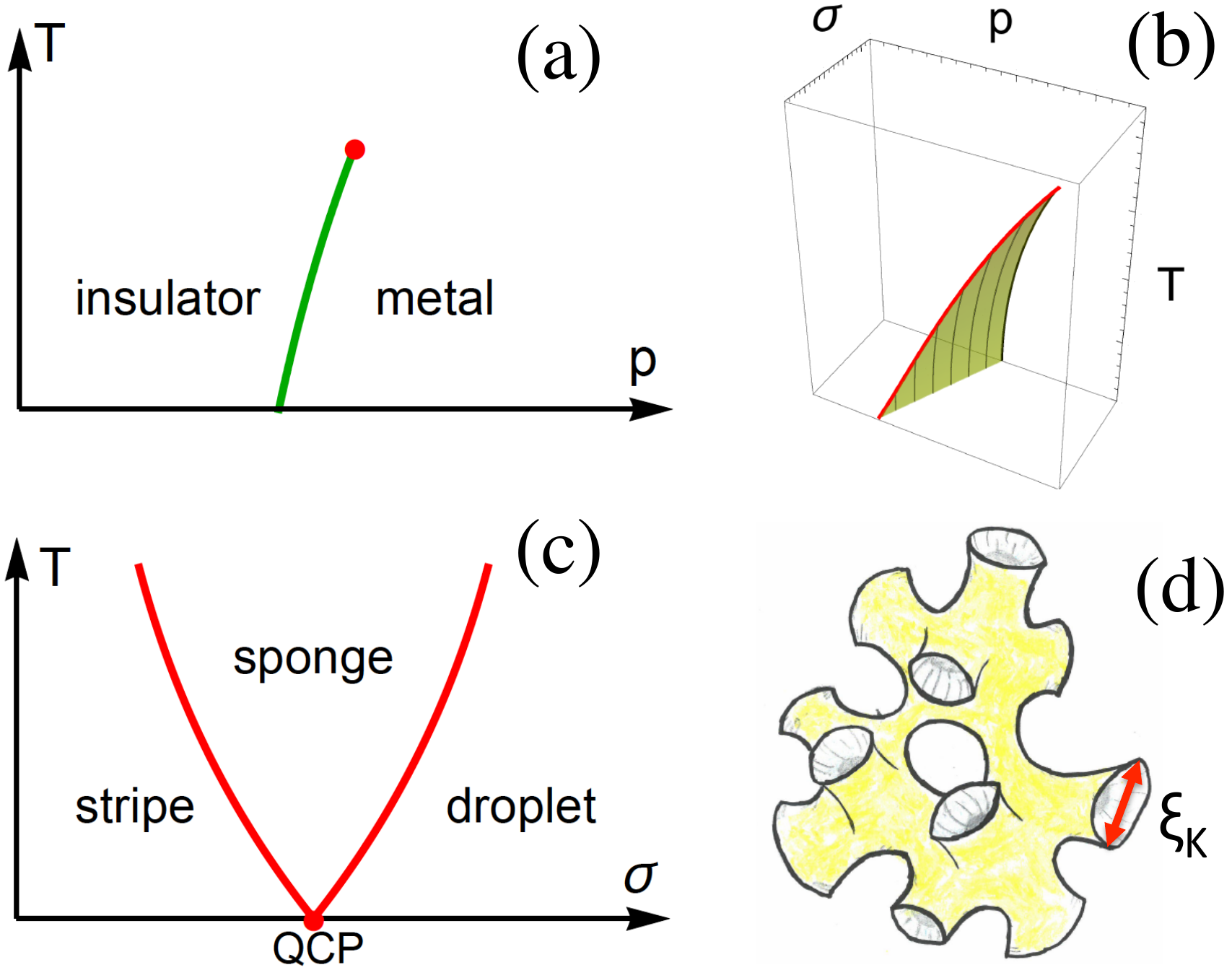}
\end{centering}
\caption{(Color online) (a) Schematic phase diagram of a first order MIT. The horizontal axis denotes the external parameter that tunes the system through the transition, the vertical axis denotes the temperature, and the red dot is the critical end point. (b) The three dimensional global phase diagram for the metal insulator transiton, including a new axis representing the interface tension $\sigma\equiv \sigma(T=0)$. As $\sigma$ decreases, the critical endpoint is driven to lower temperatures (red solid line), and becomes a QCP at $\sigma=0$ and $T=0$.  (c) The three dimensional phase diagram projected to the tension-temperature plane. The droplet phase with $\sigma>0$ corresponds to the shaded region in (b). The sponge phase is controlled by the tensionless QCP.  (d) Typical interface configuration in the sponge phase, with characteristic length scale $\xi_K$.}
\label{Fig:phase}
\end{figure}

A central object in our consideration is the interface, whose morphology gives rise to a new dimension in the global phase diagram. Let us start by considering a first order MIT (see Fig.\ref{Fig:phase}a). Close to the transition point, the system consists of mixtures of the metallic and insulating phases, separated by the interface. The length scales associated with the fluctuations of the interface are much larger than typical microscopic scales such as the interaction range and inter-carrier spacing. Such length scale separation makes it difficult to treat the interfacial degrees of freedom within a microscopic Hamiltonian. Hence we will use a coarse-grained description, where one represents the two phases as domains of ferromagnetically ordered Ising spins: spin up for the metallic phase, and spin down for the insulating phase [\onlinecite{Papanikolaou08}]. Concentrating on the behavior near the interface, one can further integrate out the bulk Ising degrees of freedom to obtain an effective Hamiltonian for the interface. In $d$-dimensions, the Hamiltonian is of the general form [\onlinecite{Helfrich73, NelsonPiranWeinberg, Chaikin}]
\begin{equation}
{\cal H}_s=\int dS \left(\sigma+\frac{\kappa}{2}H^2+{\bar \kappa} K  \right),
\label{Eq:membrane}
\end{equation}
with the $d-1$-dimensional area element $dS$, the interface tension $\sigma$, the mean curvature $H$, the Gaussian curvature $K$, and the corresponding bending rigidities $\kappa$ and ${\bar \kappa}$.

As one approaches the critical endpoint, the interface tension vanishes as $\sigma(T)\sim (1-T/T_c)^{\mu}$, with the critical exponent $\mu$ (Widom scaling). The zero temperature interface tension $\sigma(T=0)$ varies from system to system, forming a three dimensional phase diagram (Fig.\ref{Fig:phase}b). As $\sigma(T=0)$ decreases, the temperature $T_c$ of the critical endpoint decreases. When $\sigma(T=0)$ vanishes, $T_c$ also vanishes, and there is a continuous phase transiton at zero temperature, which forms a quantum critical point (QCP).
Projecting the three dimensional phase diagram to the tension-temperature plane, one obtains the phase diagram shown in Fig. \ref{Fig:phase}c. Such a phase diagram has been extensively studied in chemisty, biology and soft condensed matter physics in the context of surfactant systems [\onlinecite{Huse88, Golubovic90, Roux92}]. When the surface tension is positive and large, it is energetically favorable to minimize the surface area, and the system is in the droplet phase, with droplets of the minority phase residing in the majority phase. When the tension is negative and large, with a positive bending rigidity $\kappa$ stabilizing the system, it is energetically favorable to maximize the area of the interface, but at the same time minimize the interface curvature. The system then consists of alternating layers of metallic and insulating phases, and the system is in the lamellar (for $d=3$) or stripe (for $d=2$) phase.

The droplet phase and the lamellar/stripe phase are separated by the QCP. In the quantum critical region, the system consists of bicontinuous percolating domains of the metallic and insulating regions (see Fig.\ref{Fig:phase}d), and this phase is usually termed the sponge phase in the study of surfactant systems [\onlinecite{Huse88, Golubovic90, Roux92}]. Mapping the system to Ising spins, the droplet phase corresponds to the ferromagnetically ordered phase, and the sponge phase to the paramagnetic phase that restores the $Z_2$ symmetry. The sponge phase can also be reached by dynamically melting the stripes [\onlinecite{Zaanen01}].

\section{The model} 

We consider systems near a second-order or weakly first-order MIT, which are then in the sponge phase at finite temperatures. As a first step, we ignore the effect of disorder, which can locally pin the interface fluctuations.  Since the fermions move much faster than the interface, there is a separation of time scales. With the characteristic length scale of the sponge structure, $\xi_K$, the characteristic velocity of fermions, the Fermi velocity $v_F$, the characteristic velocity of the interface, the sound velocity $v_s$, and the condition $v_F\gg v_s$, there are two characteristic time scales: $\xi_K/v_s$ and $\xi_K/v_F$. At time scales $t<\xi_K/v_F$, assuming the fermion mean free path to be much larger than $\xi_K$, the fermion propagation is ballistic. At time scales $\xi_K/v_F<t<\xi_K/v_s$, the interface can be treated as quasi-static, and the fermions propagate in a random porous media (see Fig.~\ref{Fig:phase}d). This corresponds to the region of Knudsen diffusion [\onlinecite{Welty}], where the dominant scattering process is the scattering of fermions with the boundary. In this region, fermions are largely confined to local cages with typical size $\xi_K$, and transport occurs through tunneling between neighboring cages. At time scales $t>\xi_K/v_s$, the interface fluctuates strongly, fermions experience the averaged effect of such fluctuations, and the system is on average homogeneous.

To capture the long time scale physics, we proceed in the spirit of Hertz-Millis [\onlinecite{Hertz76, Millis93}] to describe the system in terms of fermions coupled to the bosonic collective modes, here the quantum fluctuations of the interfacial geometry. The partition function of the system involves a summation over different fermion field configurations and a summation over different geometries:
\begin{equation}
{\cal Z}=\int {\cal D}{\cal V} \int {\cal D}\psi {\cal D}\psi^\dagger e^{-{\cal S}_s[\partial{\cal V}]-{\cal S}_e[\psi, \psi^\dagger]-{\cal S}_{\rm int}[\psi, \psi^\dagger, \partial{\cal V}]},
\end{equation}
where ${\cal V}$ represents the real space configurations of the metallic regions, and $\partial{\cal V}$ their boundaries, ${\cal S}_s$ the interface action, ${\cal S}_e$ the fermion action, and ${\cal S}_{\rm int}$ the coupling between fermions and the interface. We treat the fermions, without coupling to the interface, as a renormalized Fermi gas with the action ${\cal S}_e=\int_{\cal V}d^d{\bm r}dt {\cal L}_e $, where 
\begin{equation}
{\cal L}_e=\psi^\dagger({\bm r}, t)\left(i\frac{\partial}{\partial t}+\frac{\nabla^2}{2m^*}+\mu^*\right)\psi({\bm r}, t),
\end{equation}
with effective mass $m^*$, and effective chemical potential $\mu^*$. Focusing here on the charge degrees of freedom, we have suppressed the spin index.

\subsection{Modeling the interface}

\begin{figure}[t]
\begin{centering}
\subfigure[]{
\includegraphics[width=0.8\linewidth]{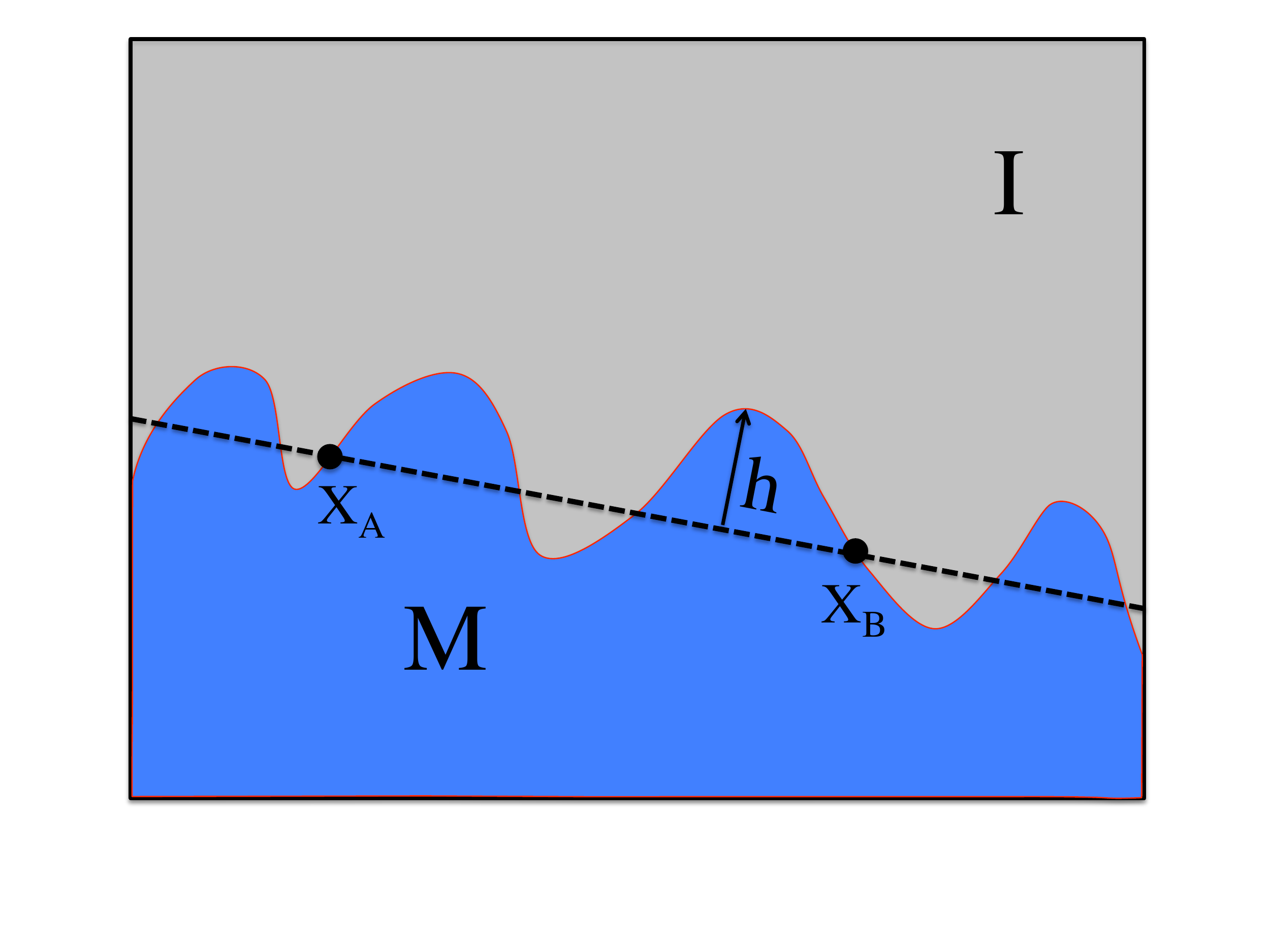} ~~}
\subfigure[]{
\includegraphics[width=0.8\linewidth]{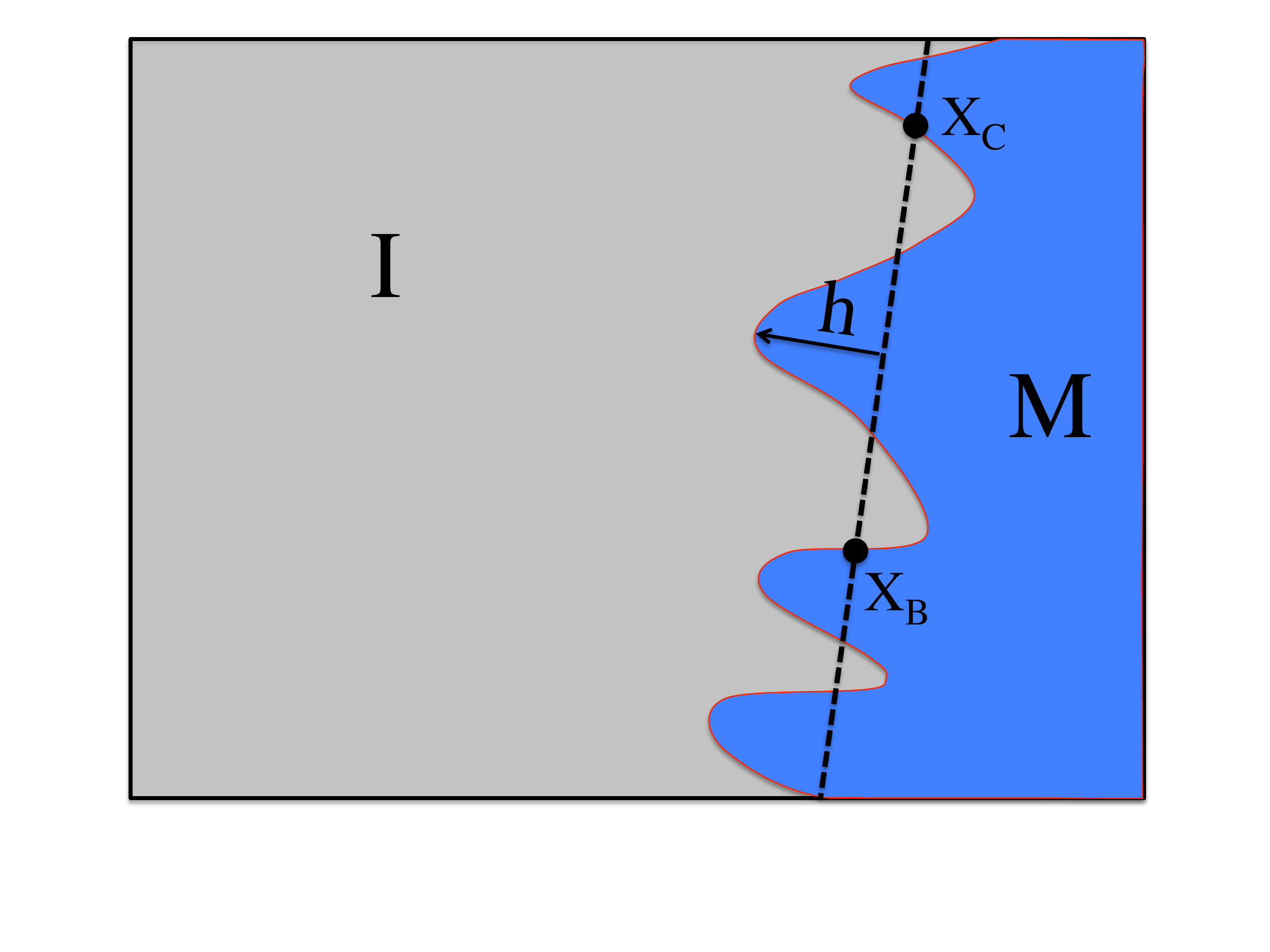} }
\end{centering}
\caption{(Color online) Typical interface configurations contributing to the bosonic string correlator $\langle {\cal O}(X_A) {\cal O}(X_B)\rangle$ (a) and $\langle {\cal O}(X_B) {\cal O}(X_C)\rangle$ (b).}
\label{Fig:cap-wave}
\end{figure}

When the interface is classical, its action takes the form ${\cal S}_s={\cal H}_s/T$. Near the MIT where quantum fluctuations are important, we also include the temporal fluctuations of the interface in the action. Let us now specify to $d=2$, for the following reasons: (1) simplicity of theoretical treatment, (2) prominent MIT materials, e.g. cuprates, organic conductors, indeed have layered structure. In this case, the fluctuating interfaces are closed bosonic strings. We use $\xi_{a}$ with $a=1, 2$ to represent the worldsheet coordinates of the string, and $X_\mu$ with $\mu=0, 1, \cdots, D-1$ its space-time coordinates. Here $D=d+1$, and in our case $d=2$. It is also useful to bear in mind that an Euclidean quantum field theory in d-dimensional space at $T=0$ can be mapped to a (d+1)-dimensional classical field theory. Hence the quantum string action in (2+1)-dimensional space-time can be mapped to the classical membrane Hamiltonian in 3-dimensions (taking $d=3$ in Eq.(\ref{Eq:membrane})).  When the string tension $\sigma$ vanishes, the string worldsheets form sponge-like structures in 2+1-dimensional space-time.

 The leading order term in the string action is the tension term, named the Nambu-Goto action [\onlinecite{Polchinski}]
\begin{eqnarray}
{\cal S}_s^{(1)}=\sigma\int d^2\xi \sqrt{g},
\end{eqnarray}
with $g\equiv\det g_{ab}$, and $g_{ab}\equiv\partial_a X^\mu\partial_bX_\mu$ is the induced metric.

Since we consider the system near a QCP associated with vanishing tension, higher order terms in the string action need to be included. Hence the string theory for interface fluctuations near the QCP goes beyond the standard textbook string theory [\onlinecite{Polchinski}] which involves only the Nambu-Goto action. The higher order terms are associated with the curvature of the worldsheet (see for example [\onlinecite{NelsonPiranWeinberg, Chaikin}] and references therein). The local curvature properties of a surface are encoded in the two principal curvatures $K_1$, $K_2$, which are the inverse of the principal curvature radii. The product of the two principal curvatures is the Gaussian curvature $K=K_1K_2$, and their average is the mean curvature $H=(K_1+K_2)/2$. 

The mean curvature term (Polyakov-Kleinert action) is of the form [\onlinecite{Polyakov86, Kleinert86}]
\begin{equation}
{\cal S}_s^{(2)}=\frac{\kappa}{2}\int d^2\xi \sqrt{g}H^2.
\end{equation}  
In terms of the string coordinates $X_\mu$, this term can be written as
\begin{equation}
{\cal S}_s^{(2)}=\frac{\kappa}{2}\int d^2\xi \sqrt{g}\left(\Delta_g X_\mu \right)^2,
\end{equation}  
where $\Delta_g X_\mu=\frac{1}{\sqrt{g}}\partial_a(\sqrt{g}g^{ab}\partial_bX_\mu )$.

The third term associated with the Gaussian curvature $K$ is the Einstein-Hilbert action of the worldsheet field theory
\begin{equation}
{\cal S}_s^{(3)}={\bar\kappa}\int d^2\xi \sqrt{g}K,
\end{equation}  
 since the Gaussian curvature $K$ is related to the scalar curvature $R$ by $K=R/2$. This term is a total divergence. The Gauss-Bonnet theorem relates such a term to the topology of the surface
\begin{equation}
\int d^2\xi \sqrt{g} K=2\pi\chi,
\end{equation} 
where $\chi$ is the Euler characteristic of the surface. Thus this term is only relevant when the topology of the worldsheet changes.

 It is quite useful to write the action in the Monge representation, $X_0=\xi_1=v_s t$, $X_1=\xi_2=x$, $X_2=h(t, x)$, where $v_s$ is the sound velocity of the interface (see Fig. \ref{Fig:cap-wave}). The area element is then $dS=d^2X\sqrt{1+(\nabla h)^2}$, with $d^2X\equiv dX_0 dX_1$, and 
\begin{equation}
\nabla h=\left( \frac{\partial h}{\partial X_0}, \frac{\partial h}{\partial X_1}\right)=\left( \frac{1}{v_s}\frac{\partial h}{\partial t}, \frac{\partial h}{\partial x}\right).
\end{equation}  
 The low energy effective action (usually named the capillary wave action, see e.g. [\onlinecite{Jasnow84}] and references therein) is obtained by expanding the action in powers of field gradients, assuming that the variation of $h$ in space-time is smooth. The Nambu-Goto action then becomes
\begin{equation}
{\cal S}_s^{(1)}= \sigma\int d^2X \left[ 1+\frac{1}{2} (\nabla h)^2\right].
\end{equation}  
 In the Monge gauge, the mean curvature reads
\begin{equation}
 H=\nabla\cdot\frac{\nabla h}{\sqrt{1+(\nabla h)^2}}.
\end{equation}
Hence the Polyakov-Kleinert term can be expanded to give
\begin{equation}
{\cal S}_s^{(2)}=\frac{\kappa}{2}\int d^2X \left( \nabla^2 h \right)^2.
\end{equation} 
The Gaussian curvature term is associated with the global topology change, and does not enter the capillary wave action. In summary, the capillary wave action reads 
$ {\cal S}_{\rm cw}=\int dx dt v_s {\cal L}_{\rm cw}$, with the Lagrangian 
\begin{eqnarray}
{\cal L}_{\rm cw}= \frac{\sigma}{2}\left[\frac{1}{v_s^2}\left( \frac{\partial h}{\partial t}\right)^2+\left( \frac{\partial h}{\partial x}\right)^2 \right] +\frac{\kappa}{2} \left( \frac{1}{v_s^2}\frac{\partial^2 h}{\partial t^2}+\frac{\partial^2 h}{\partial x^2} \right)^2.
\label{Eq:pert}
\end{eqnarray} 

\subsubsection*{Luttinger liquid perspective}
In the above consideration, the interface fluctuations are treated as noninteracting. However, since the interface fluctuations are confined to 1d, the effect of interactions can be dramatic. In fact, in 1d, both the interacting bosons and fermions are governed by Luttinger liquid physics at low energy [\onlinecite{Haldane81}]. We can use a general scaling form for the capillary wave correlator, assuming the interactions have driven the interface fluctuations to the Luttinger liquid fixed point. An equivalent way to take into account such Luttinger liquid physics is to map 
the quantum mechanically fluctuating interfaces in 2+1-d to random surfaces in 3d Euclidean space, and consider random Gaussian surfaces with a general probability distribution $e^{-{\cal S}_{\rm cw}}$, where [\onlinecite{Henley95}]
\begin{eqnarray}
{\cal S}_{\rm cw}= \frac{B}{2}\int dq d\omega (q^2+v_s^{-2}\omega^2)^{1+\zeta}|h(q, \omega)|^2,
\label{Eq:luttinger}
\end{eqnarray} 
with the stiffness $B$, and the roughness exponent $0\leq\zeta\leq 1$. One geometric measure of such random surfaces is the fractal dimension $D_f$ of the contours with equal height $h(x, t)$, which relates the length $l$ of a contour to its radius $r$, with $l\sim r^{D_f}$. The fractal dimension $1\leq D_f\leq d=2$ is related to the roughness exponent by [\onlinecite{Henley95}]
\begin{eqnarray}
D_f=\frac{3-\zeta}{2}.
\end{eqnarray} 
We note that fractal distributions of dopants have been observed in cuprates, and they are correlated with the enhancement of the superconducting transition temperature [\onlinecite{Fratini10}].

\subsection{Fermion-interface coupling}

We now specify the form of the coupling between the fermions and the interface. Similar to electron-phonon coupling in layered systems with broken inversion symmetry, e.g. in the presence of buckling [\onlinecite{Devereaux95}], we expect a local electric field arising from the asymmetric environment around the interface, with its direction perpendicular to the interface [\onlinecite{Haraldsen11, Haraldsen12}]. The interface thus possesses electric dipole charges, and the local segments of the strings can be thought of as oppositely charged parallel plates. The formation of such electric dipoles at the interface is widely observed in various heterostructures, e.g. metal-semiconductor junction, LaAlO$_3$/SrTiO$_3$ [\onlinecite{Tung01, McKee03, Nakagawa06, Goniakowski08}]. In the Monge gauge, for a static interface, the local electric field gives rise to the coupling $ {\cal S}_{\rm int}=\int dx dt v_s {\cal L}_{\rm int}$,
where
\begin{eqnarray}
{\cal L}_{\rm int}=\lambda E\psi^\dagger\psi\frac{\partial h}{\partial x}.
\end{eqnarray} 
 Consider a covariant generalization of such a coupling by introducing the antisymmetric local electromagnetic field tensor $F^{\mu\nu}$ \footnote{Since only the transverse motion of strings are relevant, one can obtain from Lorentz transformation of the electromagnetic field that $F^{\mu\nu}$ has only electric components ${\bm E}=(E_x, E_y)$, and no magnetic components.}. Here we assume the magnitude of the electric field to be fixed. The fermion density $\rho\equiv \psi^\dagger\psi$ then couples to the worldsheet operator ${\cal O}[X]=F^{\mu\nu}\epsilon^{ab}\partial_a X_\mu\partial_b X_\nu$. The interaction term thus reads
\begin{equation}
{\cal S}_{\rm int}=\lambda\int d^2\xi \psi^\dagger\psi  F^{\mu\nu}\epsilon^{ab}\partial_a X_\mu\partial_b X_\nu.
\end{equation}
The combination $\psi^\dagger\psi  F^{\mu\nu}$ plays the role of the Kalb-Ramond $2$-form field familiar in string theory [\onlinecite{Polchinski}].

\section{Fermion self-energy from interface fluctuations}

We now examine how the interface fluctuations influence the dynamics of fermions. 
The leading order corrections to the fermion action are the self-energy effect and the density-denstiy interaction. The self-energy correction is of the form
\begin{equation}
\delta S_e^{(1)}=\int d^3X_A \int d^3 X_B  \Sigma_{AB}\psi^\dagger(X_A)\psi(X_B),
\end{equation} 
with 
\begin{equation}
\Sigma_{AB}=-\lambda^2\langle {\cal O}(X_A) {\cal O}(X_B)\rangle \langle \psi(X_A) \psi^\dagger(X_B)\rangle.
\end{equation} 
The induced interaction is of the form
\begin{eqnarray}
\delta S_e^{(2)}=\int d^3X_A \int d^3 X_B  V_{AB}\rho(X_A)\rho(X_B),
\label{Eq:interaction1}
\end{eqnarray}
with the fermion density $\rho=\psi^\dagger\psi$, and 
\begin{equation}
V_{AB}=\lambda^2\langle {\cal O}(X_A) {\cal O}(X_B)\rangle.
\end{equation} 
So the effect of the interface fluctuations is encoded in the bosonic string correlator
\begin{equation}
{\cal D}_{AB}\equiv\langle {\cal O}(X_A) {\cal O}(X_B)\rangle=\int {\cal D}X{\cal O}(X_A) {\cal O}(X_B) e^{-{\cal S}_s},
\end{equation}
which involves a summation over worldsheets that pass through $X_{A, B}$, with the vertex operator ${\cal O}$ inserted at $X_{A, B}$.

The full string theoretical calculation for such a rigid string theory is beyond the scope of the present paper, and we will proceed using capillary wave theory, which captures the low energy fluctuations of the interface. 
 To calculate ${\cal D}_{AB}$, we employ the Monge gauge, where $X_0=\xi_1=v_s t$, $X_1=\xi_2=x$, $X_2=h(t, x)$. Since on average the system is homogeneous and isotropic, the coordinates can be chosen in such a way that the base plane for which $X_2=0$ passes through $X_{A, B}$. Then for given $X_{A, B}$, the correlator ${\cal D}_{AB}$ is determined from the 1+1-dimensional capillary wave field theory coupled to the 2+1-dimensional fermions,
\begin{equation}
{\cal D}_{AB}\simeq\int {\cal D}h {\cal D}\psi {\cal D}\psi^\dagger \left(E \frac{\partial h}{\partial x}\right)_A \left(E \frac{\partial h}{\partial x}\right)_B e^{- {\cal S}},
\end{equation}  
with the action $ {\cal S}={\cal S}_e+{\cal S}_{\rm cw}+{\cal S}_{\rm int}$.

We emphasize here the subtlety of dimensionality in our approach: (1) We are concerned with the long time dynamics of the system, and the system is homogeneous and isotropic on average at this time scale. Hence the correlator ${\cal D}_{AB}$ is defined in 2+1-dimensions, and depends only on the separation between $X_{A, B}$, i.e. ${\cal D}_{AB}={\cal D}(|X_A-X_B|)$. We can then Fourier transform ${\cal D}_{AB}$ from the 2+1-dimensional space-time to the momentum/frequency domain. (2) To calculate ${\cal D}_{AB}$, we need to specify the two points $X_{A, B}$, and sum over all interfaces that pass through them. We will use the capillary wave formalism for this step, that is, we sum over low energy fluctuations around the base plane that passes through $X_{A, B}$. Such capillary wave fluctuations are encoded in a 1+1-dimensional field theory, in terms of the transverse mode of the string, i.e. the height function $h(t, x)$. (3) Fermions couple to the interfaces and affect their dynamics, causing Landau damping. Since fermions live in 2+1-dimensional space-time, we will need to project their effect to the 1+1-dimensional interface.

We want to compute the Fourier transform of ${\cal D}_{AB}$ to the momentum/frequency domain, which can be obtained from the correlator ${\cal D}(r, \omega)$
\begin{equation}
{\cal D}({\bm q}, \omega)=\int d^2{\bm r} e^{-i{\bm q}\cdot{\bm r}} {\cal D}(r, \omega).
\end{equation}  
The correlator ${\cal D}(r, \omega)$, where $r\equiv |{\bm r}_A-{\bm r}_B|$, is related to the capillary wave correlator $D_h(k, \omega)\equiv 
\langle h(k, \omega) h(-k, -\omega) \rangle$ via
\begin{equation}
{\cal D}(r, \omega) = \int dk  e^{ik r}E^2k^2 D_h(k, \omega).
\end{equation}  
 After integrating over the angular coordinate, one obtains the following relation between ${\cal D}({\bm q}, \omega)$ and the capillary wave correlator
\begin{eqnarray}
{\cal D}({\bm q}, \omega)\sim \int dr r J_0(qr)\int dk  e^{ikr}E^2k^2 D_h(k, \omega),
\label{Eq:Dqw}
\end{eqnarray}
with the Bessel function $J$.  

Once ${\cal D}({\bm q}, \omega)$ is known, the one-loop fermion self-energy can be obtained from 
\begin{equation}
\Sigma({\bm k}, \omega)=\lambda^2\int d^2{\bm q} d\Omega {\cal D}({\bm q}, \Omega) G({\bm k}+{\bm q}, \omega+\Omega).
\end{equation} 
 Since the bosonic mode is much slower than the fermions, the momentum integration can be factorized into two parts $\int dq_\perp \int d q_\parallel$, with $q_\perp$ representing the momentum component transverse to the Fermi surface, i.e. the fast fermion modes, $q_\parallel$ the parallel component, i.e. the slow modes. Restricting ${\bm k}$ to be near the Fermi surface, and carrying out the $q_\perp$ integral, one obtains [\onlinecite{Chubukov05, Moon10}]
\begin{equation}
\Sigma(\omega)\sim i\lambda^2 \int d\Omega d(\Omega){\rm sign}(\omega+\Omega).
\label{Eq:Sigma}
\end{equation} 
where the bosonic correlator appears through the momentum-integrated ``local" form  
\begin{equation}
d(\omega)=\int_0^{q_0} dq_{\parallel} {\cal D}(q_\parallel, q_\perp=0, \omega),
\end{equation}  
with the momentum cutoff $q_0\sim k_F$.

We have presented above a general procedure to calculate the fermion self-energy from the 1+1-d capillary wave correlator $D_h(k, \omega)$. We now proceed to consider different models of the interface fluctuations that give specific forms of $D_h(k, \omega)$.  

\subsection{Fermion self-energy for perturbative models}

One type of models that take a more perturbative point of view is to start from Eq.(\ref{Eq:pert}), where one considers the interface to have a tension $\sigma$, a curvature stiffness $\kappa$, and a velocity $v_s$ relating frequency and momentum. Coupling to fermions induces extra dynamics for the interface fluctuations. Including such Landau damping effects, the capillary wave correlator reads (see Appendix for the calculation)
\begin{equation}
 D_h(q, \omega)=\frac{1}{\sigma(q^2+v_s^{-2}\omega^2)+\kappa(q^2+v_s^{-2}\omega^2)^2+\gamma |\omega|q^2},
\end{equation}   
with a new parameter $\gamma\propto \lambda^2$.

When the interface tension is large, the curvature term can be neglected. We consider two limiting cases. (1) When the Landau damping term is small, the dynamics is relativistic, i.e. $D_h^{-1}\simeq \sigma(q^2+v_s^{-2}\omega^2)$. The local string correlator is of the form
\begin{equation}
d(\omega)\sim \frac{v_sq_0}{\sqrt{\omega^2+v_s^2q_0^2}}.
\end{equation}  
The fermion self-energy then reads
\begin{equation}
\Sigma(\omega) \sim i {\rm ArcSinh} (\omega/v_sq_0).
\end{equation} 
After analytic continuation to real frequency, one can see that the imaginary part of the retarded self-energy ${\rm Im}\Sigma_R(\omega)=0$ for $\omega<v_sq_0$. (2) When the dynamics is controlled by Landau damping, i.e. $D_h^{-1}\simeq \sigma q^2+\gamma |\omega|q^2$, one has 
\begin{equation}
{\cal D}({\bm q}, \omega)\equiv d(\omega)\sim \frac{1}{\sigma+\gamma|\omega|}.
\end{equation}  
The self-energy then reads
\begin{equation}
\Sigma(\omega) \sim i {\rm sgn}(\omega)\log\left( 1+\frac{\gamma}{\sigma}|\omega| \right).
\end{equation} 
Analytic continuation to real frequency yields ${\rm Im}\Sigma_R(\omega)\sim \log\left(1+\frac{\gamma^2}{\sigma^2}\omega^2\right)$. At low frequencies $\omega\ll\sigma/\gamma$, ${\rm Im}\Sigma_R(\omega)\sim \omega^2$, which is of the Fermi liquid form. Hence for large interface tension, the fermions remain in the Fermi liquid phase.

When the interface tension vanishes, i.e. $\sigma\to 0$, the interface fluctuates much more violently. In this case, the dynamics is controlled by the Landau damping term, with 
\begin{equation}
D_h(q, \omega)= \frac{1}{\kappa q^4+\gamma|\omega|q^2}.
\end{equation}  
The resulting string correlator is
\begin{equation}
{\cal D}({\bm q}, \omega)\sim \frac{1}{\left(\kappa q^2+\gamma|\omega|\right)^{3/2}}.
\label{Eq:correlator}
\end{equation}
The dynamical exponent which relates the temporal direction to the spatial direction in $\omega\sim q^z$ is $z=2$.
The momentum-integrated local string correlator reads
\begin{equation}
d(\omega)\sim \frac{1}{|\omega|},
\end{equation}  
and the fermion self-energy is 
\begin{equation}
\Sigma(\omega)\sim i{\rm sgn}(\omega)\log |\omega|,
\label{Eq:Sigma}
\end{equation}  
which diverges as $\omega\to 0$. 

We can compare our result with other types of fluctuations. For example, quantum critical fluctuations associated with 2d antiferromagnetic QCPs give rise to $\Sigma(\omega)\sim \sqrt{\omega}$  [\onlinecite{Millis92, Chubukov98, Abanov01}], which is much milder than our result of a divergent self-energy. Such $\sqrt{\omega}$ form of the self-energy was also obtained in the random infinite-range interacting model [\onlinecite{Sachdev93, Parcollet99, Sachdev15}]. Similarly, quantum critical fluctuations associated with 2d ferromagnetic QCPs lead to $\Sigma(\omega)\sim \omega^{2/3}$ [\onlinecite{Rech06}].
It is surprising that the innocent-looking capillary wave field theory should have such a dramatic effect on the fermions. In fact, the capillary wave field theory can be expressed in a more familiar form by defining $\phi\equiv \partial h/\partial x$. Then the tension term is just the mass term of the bosonic field $\sigma\phi^2$, and the curvature term the gradient term $\kappa (\partial \phi)^2$. Setting $\sigma\to 0$, one obtains a massless bosonic field coupled to fermions that is well studied and no such singular effect was found. The enhancement of the self-energy correction in our case comes from the fact that the interface fluctuations reside in one dimension lower than that of the fermions. It is well known that fluctuations are more severe in lower dimensions.

\subsection{Fermion self-energy for Luttinger-liquid type models}

To put the above results in a broader perspective, let us consider another type of model that takes a more scaling point of view. From the action as shown in Eq.(\ref{Eq:luttinger}), one obtains the capillary wave correlator
\begin{equation}
D_h(q, \omega)\sim \frac{1}{(q^2+v_s^{-2}\omega^2)^{1+\zeta}}.
\end{equation}  
The resulting local string propagator is 
\begin{equation}
d(\omega)\sim \frac{1}{|\omega|^{2\zeta}},
\end{equation}  
and the fermion self-energy is of the form 
\begin{equation}
\Sigma(\omega)\sim i{\rm sgn}(\omega)|\omega|^{1-2\zeta}.
\label{Eq:Sigma2}
\end{equation} 
The self-energy diverges for rough surfaces with $\zeta\geq 1/2$. For $\zeta=1/2$, where $d(\omega)\sim 1/|\omega|$, the self-energy is $\Sigma(\omega)\sim i{\rm sgn}(\omega)\log|\omega|$, which diverges logarithmically. This case corresponds to the above perturbative model with vanishing tension $\sigma=0$.

We can take an even more coarse-grained scaling point of view to regard the local string propagator $d(\omega)$ as having a certain scaling form, and use that as input for the calculation of the fermion self-energy. It is interesting to note that the marginal fermi liquid with $\Sigma(\omega)\sim i\omega\log(\omega_0/\omega)$ corresponds to $d(\omega)\sim \log|\omega|$, i.e. $\zeta\to 0$.

\subsection{Zero of the Green's function}

We found above that the interface fluctuations can give rise to a divergent fermion self-energy. The fermion Green's function $G({\bm k}, \omega)=\frac{1}{i\omega-\xi_{\bm k}-\Sigma(\omega)}$ is then dominated by the self-energy term at low frequencies. For the perturbative model with $\sigma=0$, one has
\begin{equation}
G({\bm k}, \omega)\sim \frac{i{\rm sgn}(\omega)}{\log|\omega|}.
\end{equation}  
For the Luttinger-liquid type model with $\zeta>1/2$, one has
\begin{equation}
G({\bm k}, \omega)\sim i{\rm sgn}(\omega)|\omega|^{2\zeta-1}.
\end{equation}  
Due to the singular frequency dependence, the momentum dependence becomes negligible, and the full fermion Green's function becomes local. Furthermore, the original pole of free Fermi gas Green's function at $\omega=0$ is converted to a zero by the divergent self-energy, which indicates that propagating single electron states cease to exist. The Landau Fermi liquid picture completely breaks down, and we obtain a non-Fermi liquid, i.e. strange metal state. The absence of propagating single particle states in our system is intuitively clear: at the critical point, where the interface tension vanishes, the interface forms sponge-like structures at large scales (see Fig.\ref{Fig:phase}(d)). Hence the electrons scatter so much with the interfaces on their way of propagation that they can no longer propagate at large spatial/temporal scales, and are effective confined.

 It is interesting to note that, the conversion of poles of the fermion propagator to zeros also occurs in the $1+1$-d large-$N$ gauge theory, which is a toy model for quark confinement [\onlinecite{tHooft74}]. There such behavior is regarded as a clear signature of the absence of physical single quark states. The significance of the zeros of the Green's function and in particular their relation with the Luttinger theorem have been discussed in [\onlinecite{Dzyaloshinskii03, Phillips13}].

With the absence of propagating single particle states at low energies, the resistivity of such non-Fermi liquid metals is no longer controlled by the quasiparticle decay rate as in a Fermi liquid metal, but by the decay rate of the total momentum. Since the system equilibrates very rapidly, at scales shorter than that of the breaking of momentum conservation due to, for example, impurity and Umklapp scattering, the transport properties of the system can be described by hydrodynamics [\onlinecite{Spivak11, Zaanen14}]. The temperature dependence of resistivity is then determined by the scaling dimension of energy density and charge density operators [\onlinecite{Hartnoll12, Zaanen14}]. One way to compute the resistivity for systems near the MIT is to employ the random resistor network [\onlinecite{Papanikolaou08}]. The quantum-critical sponge phase identified here corresponds to the spin liquid phase of a frustrated Ising model. It would be interesting to calculate the resistivity of the resistor network defined on a geometrically frustrated lattice. We leave this for a future investigation.

\section{Interface fluctuation mediated pairing}

The interface fluctuations induce a density-density interaction among the fermions as shown in Eq.(\ref{Eq:interaction1}). It can be written more explicitly as 
\begin{eqnarray}
\delta {\cal S}_e^{(2)}\sim &&-\lambda^2 \int dt_1 \int d^2{\bm r}_1 \int dt_2  \int d^2{\bm r}_2 {\cal D}({\bm r}, t)\nonumber\\
&& \psi^\dagger_\alpha({\bm r}_1, t_1)\psi_\alpha({\bm r}_1, t_1) \psi^\dagger_\beta({\bm r}_2, t_2)\psi_\beta({\bm r}_2, t_2),
\end{eqnarray} 
with the bosonic string correlator ${\cal D}$, and ${\bm r}={\bm r}_1-{\bm r}_2$, $t=t_1-t_2$. Here we have restored the spin indices $\alpha, \beta=\uparrow, \downarrow$, and the summation over them is kept implicit to simplify the notation.

 As is familiar from the case of phonon mediated pairing, such density-density interaction can give rise to Cooper pairing in the spin singlet channel. We note the difference with models employing homogeneous approaches [\onlinecite{Lee06, Mross11}], where a repulsive current-current interaction is obtained between two fermions with opposite momentum that pair [\onlinecite{Metlitski15}]. In our case, the local electric field resulting from inversion symmetry breaking around the interface induces a direct coupling between the fermion density and the bosonic mode, which leads to an attractive density-density interaction among the fermions. However, as shown above, the interface fluctuations also generate self-energy corrections for fermions, which lead to incoherence, and are actually detrimental for superconductivity. Indeed for $\sigma=0$ in the perturbative models, and for $\zeta \geq 1/2$ for the Luttinger-liquid type models, the self-energy diverges. The pole in the single electron Green's function is converted to a zero, and there is no physical single electron state. However, the existence of Cooper pairs is associated with the pole of the two electron Green's function in the particle-particle channel. To see whether such a pole exists at the one-loop level, we can use the Eliashberg formalism, which takes into account the competing effects of self-energy and density-density interaction. The Eliashberg formalism has been applied to systems of fermions coupled to quantum critical fluctuations in [\onlinecite{Abanov01, Chubukov05, Moon10, She11, Chung13}].

The density-density interaction can be decomposed in the pairing channel. Keeping terms relevant for zero total-momentum spin-singlet pairing, one has 
\begin{eqnarray}
\delta {\cal S}_e^{(2)}\sim &&-\lambda^2 \int d^2{\bm k} \int d\omega \int d^2{\bm k}' \int d\omega' {\cal D}({\bm k}-{\bm k}', \omega-\omega')\nonumber\\ &&\chi^\dagger({\bm k}, \omega)\chi({\bm k}', \omega'),
\end{eqnarray} 
with the spin-singlet pairing operator $\chi({\bm k}, \omega)\equiv \frac{1}{\sqrt{2}}(\psi_{{\bm k} \omega \uparrow}\psi_{-{\bm k}, -\omega, \downarrow}-\psi_{{\bm k} \omega \downarrow}\psi_{-{\bm k}, -\omega, \uparrow})$. The superconducting gap can then be defined from
\begin{eqnarray}
\Delta({\bm k}, \omega_n)\equiv\lambda^2 T\sum_{{\bm k}', \omega'_m}{\cal D}({\bm k}-{\bm k}', \omega_n-\omega'_m)\langle\chi({\bm k}', \omega'_m)\rangle.
\end{eqnarray} 
 Since the interface fluctuation mediated interaction is isotropic, pairing is in the $s$-wave channel. Other degrees of freedom, e.g. spin fluctuations [\onlinecite{Monthoux91}], need to be invoked to generate unconventional, e.g. $d$-wave, pairing.

\begin{figure}
\begin{centering}
\includegraphics[width=0.9\linewidth]{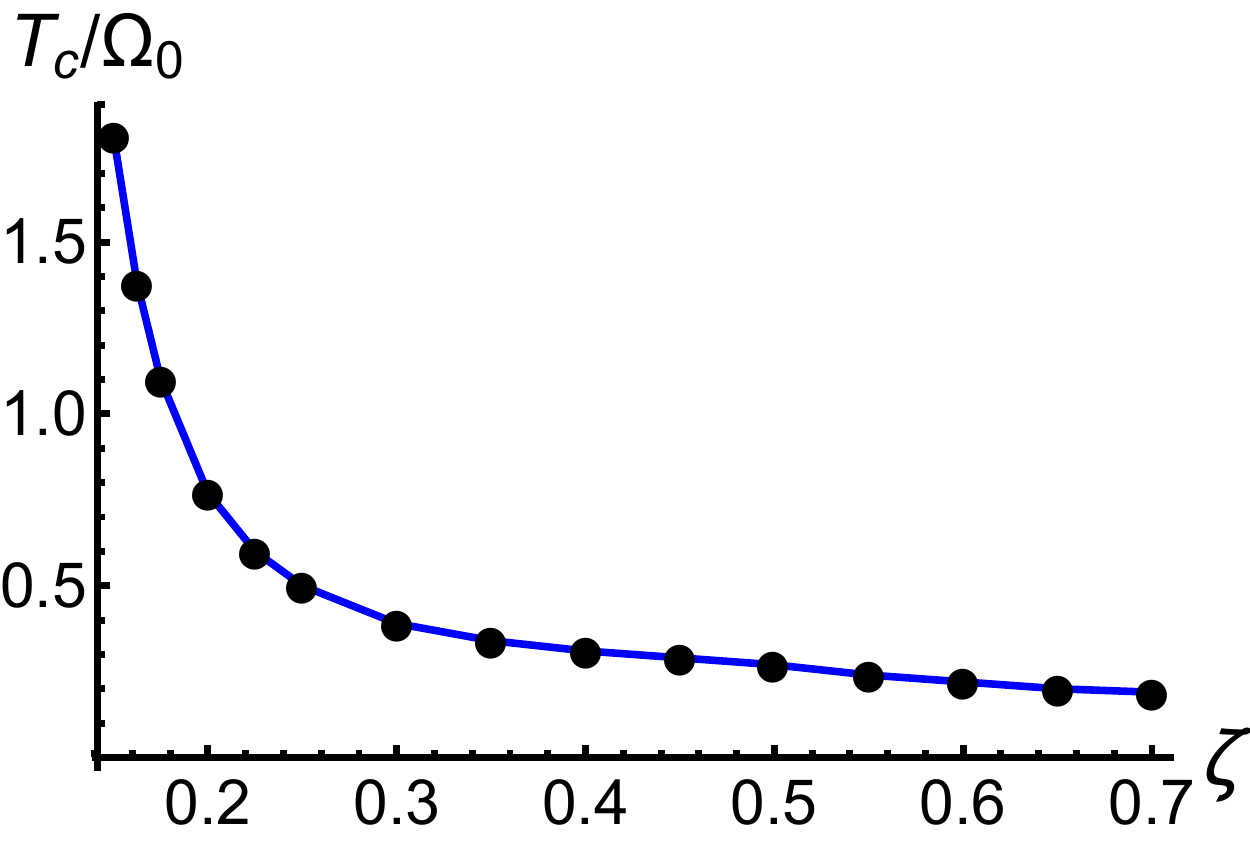}
\end{centering}
\caption{(Color online) The pairing temperature as function of the critical exponent $\zeta$ in the bosonic propagator.}
\label{Fig:Tc}
\end{figure}

With the fermions moving much faster than the bosons, momentum can integrated out in the Eliashberg formalism, leaving only the frequency dependence explicit. Near the superconducting transition, one has the coupled equations for the self-energy $\Sigma(\omega_n)$ (Eq.\ref{Eq:Sigma}, \ref{Eq:Sigma2}) and the pairing vertex $\Phi(\omega_n)$ (see. e.g. [\onlinecite{Moon10}])
\begin{eqnarray}
\Phi(\omega_n)=\pi T\sum_m \frac{d(\omega_m-\omega_n)}{|\omega_m|Z(\omega_m)}\Phi(\omega_m),
\label{Eq:Phi}
\end{eqnarray} 
with the pairing vertex  $\Phi(\omega_n)=\Delta(\omega_n)Z(\omega_n)$, and the renormalization factor $Z(\omega_n)=1+\Sigma(\omega_n)/\omega_n$. Pairing occurs when the gap equation (Eq.(\ref{Eq:Phi})) have a solution. In particular, for a quantum critical boson with the local correlator $d(\omega)=(\Omega_0/|\omega|)^{2\zeta}$, where $0<\zeta <1$, Eq.(\ref{Eq:Phi}) always has a solution.  Since $\Omega_0$ is the only dimensionful parameter here, the resulting pairing temperature $T_c$ is set by $\Omega_0$, with [\onlinecite{Abanov01, Chubukov05, Moon10}]
\begin{eqnarray}
T_c=\Omega_0 f(\zeta).
\end{eqnarray} 
The coefficient $f(\zeta)$ decreases with increasing $\zeta$ (see Fig.\ref{Fig:Tc}), implying a  stronger effect of incoherence.

The above calculations imply a novel picture for strange metals: while there are no propagating single particle excitations, there can be Cooper pairs. Intuitively one expects fermions confined in the sponge structures to be unstable, since their zero point energy is significantly boosted by the confinement. Pairing is a plausible way to release such energy. In fact, superconductivity has been widely observed near Fermi-surface-changing MIT in materials like cuprates [\onlinecite{Imada98, Lee06}], organic conductors [\onlinecite{Powell11, Ardavan12}], iron pnictides [\onlinecite{Paglione10, Shibauchi14}] and heavy fermions [\onlinecite{Stewart01, Lohneysen07}]. However one needs to be cautious with the above perturbative treatment of pairing. We discuss below a more local picture of pairing in the fluctuating sponge structure.

At time scales where the interface can be treated as quasistatic, the electrons are localized to cells of size $\xi_K$ due to the inhomogeneous background. Then Cooper pairing occurs locally. In 2d the binding energy of two electrons scales with the size of the system as $E_b\sim 1/(L^2\log L)$ (see Appendix II). With $L\sim \xi_K$, pairing is largely enhanced by confining electrons in the sponge structures. This effect is in some sense the real space counterpart of Cooper's mechanism of pairing [\onlinecite{Cooper56}]. The celebrated Cooper instability relies on the existence of the Fermi surface, and electron binding is enhanced due to Pauli blocking in momentum space. In our case, electron binding is enhanced due to the temporary confinement of electrons in real space. The occurence of superconductivity also requires phase coherence. In the fluctuating sponge strucutre, electrons can first pair locally, taking advantage of the temporary confinement of the local geometry. And then, at larger time scales, where the geometric fluctuations come into action, the whole system becomes phase coherent. Hence pairing and phase coherence have different time scales, and correspondingly different energy scales. A more in-depth investigation of this pairing mechanism is left for a future publication.

\section{Thermodynamic instability}

\begin{figure}
\begin{centering}
\includegraphics[width=0.9\linewidth]{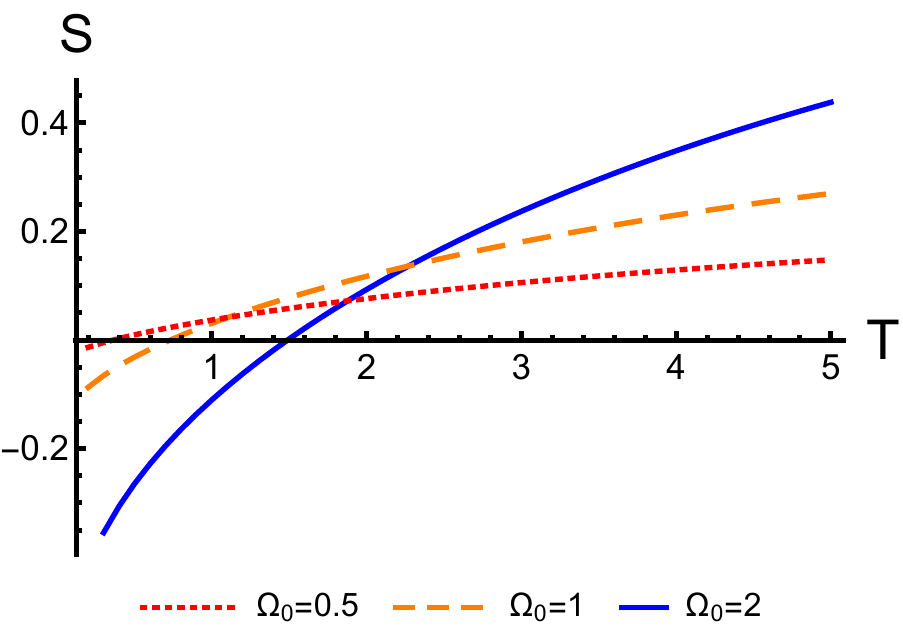}
\end{centering}
\caption{(Color online) The entropy as function of temperature for three different characteristic interaction scales.}
\label{Fig:entropy}
\end{figure}

We have shown above that the interface fluctuations induce pairing instability among the electrons. Here we will show that there is indeed a deeper reason for the occurrence of pairing: the strange metal phase is thermodynamically unstable below certain temperature $T^*$, and hence some new phase must set in at such a temperature. The corresponding temperature scale is on the order of the characteristic interaction energy scale $\Omega_0$, which, as shown above, also determines the pairing scale, i.e. $T^*\sim T_c\sim \Omega_0$.

The entropy of the system can be determined from the relation [\onlinecite{AGD}]    
\begin{eqnarray}
\frac{S}{V}=\frac{-1}{\pi i T}\int \frac{d^2{\bm k}}{(2\pi)^2} \int d\omega \omega\frac{\partial f(\omega)}{\partial \omega} 
\left[\log G^R_{{\bm k}, \omega}-\log G^A_{{\bm k}, \omega} \right],
\end{eqnarray} 
where $V$ is the volume of the system, and $f$ is the Fermi function. $G^R_{{\bm k}, \omega}$ and $G^A_{{\bm k}, \omega}$ are the retarded and advanced fermion Green's functions, and are obtained from the Matsubara Green's function through analytic continuation. For the self-energy of the form as shown in Eq.(\ref{Eq:Sigma}), they read
\begin{eqnarray}
 G^{R,A}_{{\bm k}, \omega}=\frac{1}{\omega-\xi_{\bm k}-\frac{\pi}{2}\Omega_0 {\rm sgn}(\omega)\pm i \Omega_0\log\frac{\Lambda}{|\omega|}},
\end{eqnarray}   
with the characteristic interaction scale $\Omega_0$, and the cutoff $\Lambda$.

The results are shown in Fig.~\ref{Fig:entropy}. We find that the entropy decreases with lowering temperature, and it becomes negative at $T^*=A\Omega_0$, with $A$ of order unity (here $A\simeq 0.73$). Since $\Omega_0$ is the only dimensionful parameter here (besides the cutoff), it is not surprising that $T^*$ is on the order of $\Omega_0$. The negative entropy implies that the strange metal phase becomes thermodynamically unstable for temperatures $T<T^*\sim \Omega_0$. Indeed, as shown above, pairing sets in at the same temperature scale $T_c\sim \Omega_0$. The concurrence of the two temperature scales $T^*\sim T_c$ originates from the fact both the self-energy and the pairing interaction have the same characteristic energy scale $\Omega_0\sim \lambda^2 \int {\cal D}$, which is determined by the coupling strength $\lambda$ and the bosonic propagator ${\cal D}$. The resulting picture is that the strange metal phase is intrinsically unstable, and it is forced thermodynamically to be replaced by a symmery-broken phase at low temperatures.

\section{Conclusions}

 We have proposed a new framework for non-Fermi liquid metals near a Mott metal-insulator-transition with fermions operating in sponge-like structures formed by strongly fluctuating interfaces. The geometric fluctuations of the interface give rise to anomalous scalings in the fermion self-energy that are qualitatively different from those of Landau Fermi liquids. Furthermore the geometric fluctuations induce pairing instability in the absence of external bosonic glues. The resulting picture of a strange metal phase at high temperatures and a low temperature superconducting phase is in qualitative agreement with the experimental phase diagram of cuprates, iron pnictides, organic conductors and heavy fermions.

By directly incorporating quantum geometric fluctuations into our framework, our result constitutes a qualitatively new step going beyond the homogeneous paradigm of many body systems. The fluctuating geometry is not a perturbation to the system, but instead provides the template for all other degrees of freedom. It will be interesting to study other processes, e.g. spin-fluctuation mediated pairing [\onlinecite{Scalapino12}], in such a fluctuating background.

A crucial question is how to distinguish between the effects of geometric fluctuations and fluctuations of other degrees of freedom. One strategy is to look for strange metal behavior away from the symmetry-breaking QCPs. In fact, strange metal behavior was observed in Ir- and Ge-substituted YbRh$_2$Si$_2$ [\onlinecite{Friedemann09, Custers10}] and $\beta$-YbAlB$_4$ under pressure [\onlinecite{Tomita15}] in regions associated with a Mott transition away from the magnetic QCPs. Quantum geometric fluctuations may play important roles in these systems.

One further extension of our work is to consider fluctuating interfaces involving topologically nontrivial phases, for which there can be gapless fermion modes emerging at the interface. A concrete example is a heterostructure of a 3d topological insulator with a magnetic insulator near the ferromagnetic transition. When approaching the transtion from the ordered side, magnetic domains will be created. A gapless chiral fermion mode is then induced on the topological insulator at the location of the domain wall [\onlinecite{Kane10, Qi11}]. The domain walls are then heterotic strings [\onlinecite{Gross85}] in the sense that the fermionic mode is chiral while the bosonic modes have both left- and the right-moving components.

\section*{Acknowledgements}
We are grateful to Steve Kivelson, Liam McAllister, Erich Mueller, Subir Sachdev, Yu-dai Tsai and Jan Zaanen for useful discussions. Work at Los Alamos was supported by DOE BES E 304. Work at Nordita was supported by ERC DM 32103 and KAW. Work at Cornell was supported by the Cornell Center for Materials Research with funding from the NSF MRSEC program (DMR-1120296).

\section*{Appendix I: Landau damping}

Coupling to fermions leads to Landau damping in the capillary wave correlator. One has the Dyson equation
\begin{equation}
 D_h^{-1}=[ D_h^{(0)}]^{-1}+\Pi_h,
\end{equation}  
 with the free part 
\begin{equation}
[ D_h^{(0)}]^{-1}=\sigma(k^2+v_s^{-2}\omega^2)+\kappa(k^2+v_s^{-2}\omega^2)^2,
\end{equation} 
 and the self energy correction $\Pi_h$ coming from coupling to fermions. The 2+1-dimensional fermion bubble 
\begin{equation}
\Pi_0({\bm q}, \Omega)\equiv\int\frac{d^2{\bm k}}{(2\pi)^2}\frac{d\omega}{2\pi}G_0({\bm k}, \omega)G_0({\bm k}+{\bm q}, \omega+\Omega)
\end{equation} 
 is of the form 
\begin{equation}
\Pi_0({\bm q}, \Omega)\sim |\Omega|/q.
\end{equation} 
 Fourier transforming to 2+1-dimensional real space yields 
\begin{equation}
\Pi_0({\bm r}, \Omega)=\int\frac{d^2{\bm q}}{(2\pi)^2} e^{i{\bm q}\cdot{\bm r}}\Pi_0({\bm q}, \Omega)\sim \frac{|\Omega|}{r}.
\end{equation}
 Since the system is homogeneous and isotropic on average, $\Pi_0({\bm r}, \Omega)=\Pi_0(r, \Omega)$ determines the self energy in the 1+1-dimensional field theory. With the fermion density coupling directly to $\partial h$, one obtains the self energy correction for $\langle\partial h\partial h\rangle$ as 
\begin{equation}
\Pi_{\partial h}(k, \omega)=\lambda^2E^2\int dr e^{-ikr}\Pi_0(r, \omega)\sim -(\gamma_0+\log|k|)|\Omega|,
\end{equation} 
with Euler's constant $\gamma_0$. We note that there is a logarithmic divergence at small momenta, and an IR cutoff needs to be imposed. The self energy correction for the $\langle h h\rangle$ correlator is then 
\begin{equation}
\Pi_h(k, \omega)=k^2\Pi_{\partial h}(k, \omega).
\end{equation} 
Keeping only the leading power law dependence, the result is of the form 
\begin{equation}
\Pi_h(k, \omega)=\gamma |\omega|k^2,
\end{equation} 
 with the prefactor $\gamma\propto \lambda^2E^2$. We need to keep in mind that $\gamma$ depends on the IR cutoff. The capillary wave correlator thus reads
\begin{equation}
 D_h(k, \omega)=\frac{1}{\sigma(k^2+v_s^{-2}\omega^2)+\kappa(k^2+v_s^{-2}\omega^2)^2+\gamma |\omega|k^2}.
\end{equation}  

We also note that with the full Green's function $G({\bm k}, \omega)\sim i{\rm sgn}(\omega)/\log|\omega|$, the fermion bubble gives
\begin{equation}
\Pi_0({\bm q}, \Omega)\sim \int d\omega \frac{{\rm sgn}(\omega)}{\log|\omega|}\frac{{\rm sgn}(\omega+\Omega)}{\log|\omega+\Omega|}\sim |\Omega|.
\end{equation} 
The leading power dependence of $\Pi_h(k, \omega)$ is then the same as using the free fermion Green's function, and the above form of $ D_h(k, \omega)$ is self-consistent.

\section*{Appendix II: Cooper pairing in a confining geometry}

Consider two particles in a harmonic trap interacting via a pointlike force with Hamiltonian $H=H_0+H_1$, where [\onlinecite{Busch98}]
\begin{eqnarray}
H_0&=&-\frac{\hbar^2}{2m}\nabla_1^2-\frac{\hbar^2}{2m}\nabla_2^2+4\pi \frac{\hbar^2}{m}a_0\delta_{\rm reg}({\bm r}_1-{\bm r}_2),\\
H_1&=&\frac{1}{2}m\omega_0^2r_1^2+\frac{1}{2}m\omega_0^2r_2^2,
\end{eqnarray}
with the scattering length $a_0$, and the regularized $\delta$ function potential $\delta_{\rm reg}({\bm r})$. Such a point-like potential is a good approximation for $a_0\lesssim L$, where $L=\sqrt{\hbar/(m\omega_0)}$ is the characteristic length scale of the harmonic trap. The characteristic energy scale of the harmonic trap is $E_L=\hbar \omega_0=\hbar^2/(mL^2)$.

In 3d, the binding energy is related to the size of the trap via
\begin{equation}
\sqrt{2}\frac{\Gamma\left(\frac{E_b}{2E_L}\right)}{\Gamma\left(\frac{E_b}{2E_L}-\frac{1}{2}\right)}=\frac{L}{a_0}.
\end{equation}
When the trap size is on the order of the scattering length, $L\sim a_0$, the binding energy is $E_b\sim \hbar^2/(m a_0^2)$. $E_b$ increases rapidly as $L$ decreases, and decays rapidly as $L$ increases. Mathematically bound states always exist in the presence of a trap. But practically the binding energy is too small to lead to any observal effect for $L\gg a_0$. At large $L$, one has 
\begin{equation}
E_b\sim \frac{\hbar^2a_0}{mL^3}.
\end{equation}

 In 2d, the binding energy is given by [\onlinecite{Busch98}]
\begin{equation}
\psi\left(\frac{E_b}{2E_L}\right)=\log\left(\frac{L^2}{2a_0^2} \right), 
\end{equation}
where $\psi$ is the digamma function. For $x\to 0$, $\psi(x)\simeq -\gamma-1/z$. At large $L$, one has 
\begin{equation}
E_b\sim \frac{\hbar^2}{mL^2\log(L^2/a_0^2)}.
\end{equation}

\bibliographystyle{apsrev}
\bibliography{strings,refs}

\end{document}